\documentclass[12pt]{iopart}

\usepackage{amssymb}
\usepackage{graphicx}
\usepackage{dcolumn}
\usepackage{bm}
\usepackage{psfrag}
\usepackage{bbold}
\usepackage{amsthm}
\usepackage{iopams}

\newcommand{\ran}{\right\rangle}
\newcommand{\ket}[1]{\left| #1 \ran}

\newcommand{\av}[1]{\left\langle  #1  \right\rangle}

\usepackage{xcolor}

\newcommand{\reff}[1]{(\ref{#1})}
\newcommand{\mal}{\mathcal}

\newcommand{\abs}[1]{\left| #1 \right|}

\newcommand{\lt}{\left(}
\newcommand{\rt}{\right)}

\begin{document}

\title{Non-equilibrium universality in the dynamics of dissipative cold atomic gases}

\author{M Marcuzzi, E Levi, W Li, J P Garrahan, B Olmos and I Lesanovsky}

\address{School of Physics and Astronomy, University of Nottingham, Nottingham, NG7 2RD, UK}
\ead{beatriz.olmos-sanchez@nottingham.ac.uk}

\begin{abstract}
The theory of continuous phase transitions predicts the universal collective properties of a physical system near a critical point, which for instance manifest in characteristic power-law behaviours of physical observables. The well-established concept at or near equilibrium, universality, can also characterize the physics of systems out of equilibrium. The most fundamental instance of a genuine non-equilibrium phase transition is the directed percolation universality class, where a system switches from an absorbing inactive to a fluctuating active phase. Despite being known for several decades it has been challenging to find experimental systems that manifest this transition. Here we show theoretically that signatures of the directed percolation universality class can be observed in an atomic system with long range interactions. Moreover, we demonstrate that even mesoscopic ensembles --- which are currently studied experimentally --- are sufficient to observe traces of this non-equilibrium phase transition in one, two and three dimensions.
 \end{abstract}


\maketitle

\section{Introduction}

In strongly interacting systems, long range correlations can lead to the emergence of collective behaviours which can be distinctively different from the single-body physics governing the constituents. A remarkable example can be found within the theory of continuous phase transitions \cite{Ma, Goldenfeld_book, Zinn-Justin}, which indeed predicts characteristic singular behaviours of macroscopic observables, in contrast to the usually smooth microscopic description. At a so-called critical point, the correlation length of the system diverges and the overall behaviour is fundamentally determined by certain properties --- such as dimensionality, range of interactions and symmetries --- that do not depend on the specific scale. All systems sharing these few coarse-grained features display the same qualitative macroscopic physics \cite{Exp_univ}
and form a \emph{universality class}, which is in turn characterized by the corresponding \emph{critical exponents and ratios} \cite{Ma, Goldenfeld_book, Zinn-Justin, PelVicari}.

Phase transitions have been extensively studied in equilibrium --- both in the classical regime \cite{Huang, Goldenfeld_book, Zinn-Justin, LeBellac} and in the quantum one \cite{QPT, Mussardo} --- within the general framework of statistical mechanics, based on the properties of a few macroscopic thermodynamic potentials. Out of equilibrium, the lack of an equivalent unified picture has prevented reaching a similar degree of insight. Despite this, critical phenomena are known to occur out of equilibrium as well, e.g., in the statistics of the work \cite{Work_GS1, Work_GS2, Work_GS3}, the temporal evolution \cite{DPT1, DPT2, DPT3, DPT4} and the spectral properties \cite{DPT_Diehl} of quenched quantum systems. In particular, phase transitions can take place in the properties of the steady state; this typically leads to an enrichment of the stationary phase diagram which depends upon the coarse-grained aspects of the dynamics, such as symmetries and conservation laws (see, e.g., \cite{HH,Driven1}). Equilibrium conditions, for example, are specifically related to the \emph{microreversibility} symmetry \cite{Microrev1, Microrev2, Microrev3, Chiocchetta15}. Systems in which the latter is not recovered at large length and time scales undergo genuine non-equilibrium phase transitions \cite{NEQ_PT1, DP_Hinrichsen, NEQ_Lattice}.

One of the most fundamental and well-studied universality classes of this kind is \emph{directed percolation} (DP). A DP-like transition is defined as a continuous transition between a \emph{unique absorbing state} and a fluctuating one observed for a \emph{positive, one-component order parameter}. An intuitive description can be provided in terms of the ``contact process'', i.e., a classical stochastic process on a chain of Ising spins in which an up spin (or excitation) can always flip down (self-destruction), whereas a down spin can only flip up if another up spin is present in its neighbourhood (offspring production), see Fig.~\ref{fig:fig2}. Due to the fact that the latter of these processes require a nearby up spin to take place, the ``all-down'' state constitutes not only a stationary state of the dynamics, but an absorbing one as well (i.e. once the system is in this state it cannot abandon it), being completely free of stochastic fluctuations. When the rate at which self-destruction occurs is much larger than the rate of offspring production, the system invariably ends up in the absorbing state, no matter what the initial conditions were. At a critical ratio between the rates, the system switches continuously from the ``all-down'' steady state to a fluctuating active one with a finite density of excitations. Despite its simplicity and robustness \cite{DP_Hinrichsen}, it has been very difficult to identify clear signs of DP universality in physical systems \cite{Hinrichsen_exp}. Only recently an experiment focussing on two distinct topological phases of nematic liquid crystals provided the first clean realization of DP in a two-dimensional physical system \cite{DP_exp, DP_explong}.


In this work we show that strongly interacting ensembles of highly excited (so-called Rydberg) atoms \cite{Bloch,Rydberg2,Ryd-QI} feature a dynamics which --- in a certain limit --- is governed by the elementary rules of a DP process \cite{DP_Hinrichsen}. Beyond revealing insights into the out-of-equilibrium behaviour of this currently much studied system, our work highlights an alternative approach for the experimental exploration of DP universal features not only in two but in one and three dimensions as well. Remarkably, our results show that even for relatively small (mesoscopic) system sizes which compare to those currently studied in experiment \cite{Viteau11,Ryd-lattice2,Schauss14,Barredo14}, clear signatures of DP are observable.

This paper is structured as follows. In Section \ref{sec:Setup} we briefly introduce the open quantum many-body system under consideration and focus on a parameter regime where its dynamics can be described via a classical rate equation. Section \ref{sec:DP} shows how the parameters of the system can be tuned appropriately so that this dynamics is very close to that of a DP process, which is then further tested at the mean-field level in Section \ref{sec:MF}. Numerical simulations of mesoscopic systems in one and two dimensions indicate that indeed strong signatures of the DP universality class can be observed in this long range interacting system (Section \ref{sec:Numerics}). Our concluding remarks can be found in Section \ref{sec:Conclusions}.

\section{The setup}\label{sec:Setup}

\begin{figure}
\center
  \includegraphics[width=0.7\columnwidth]{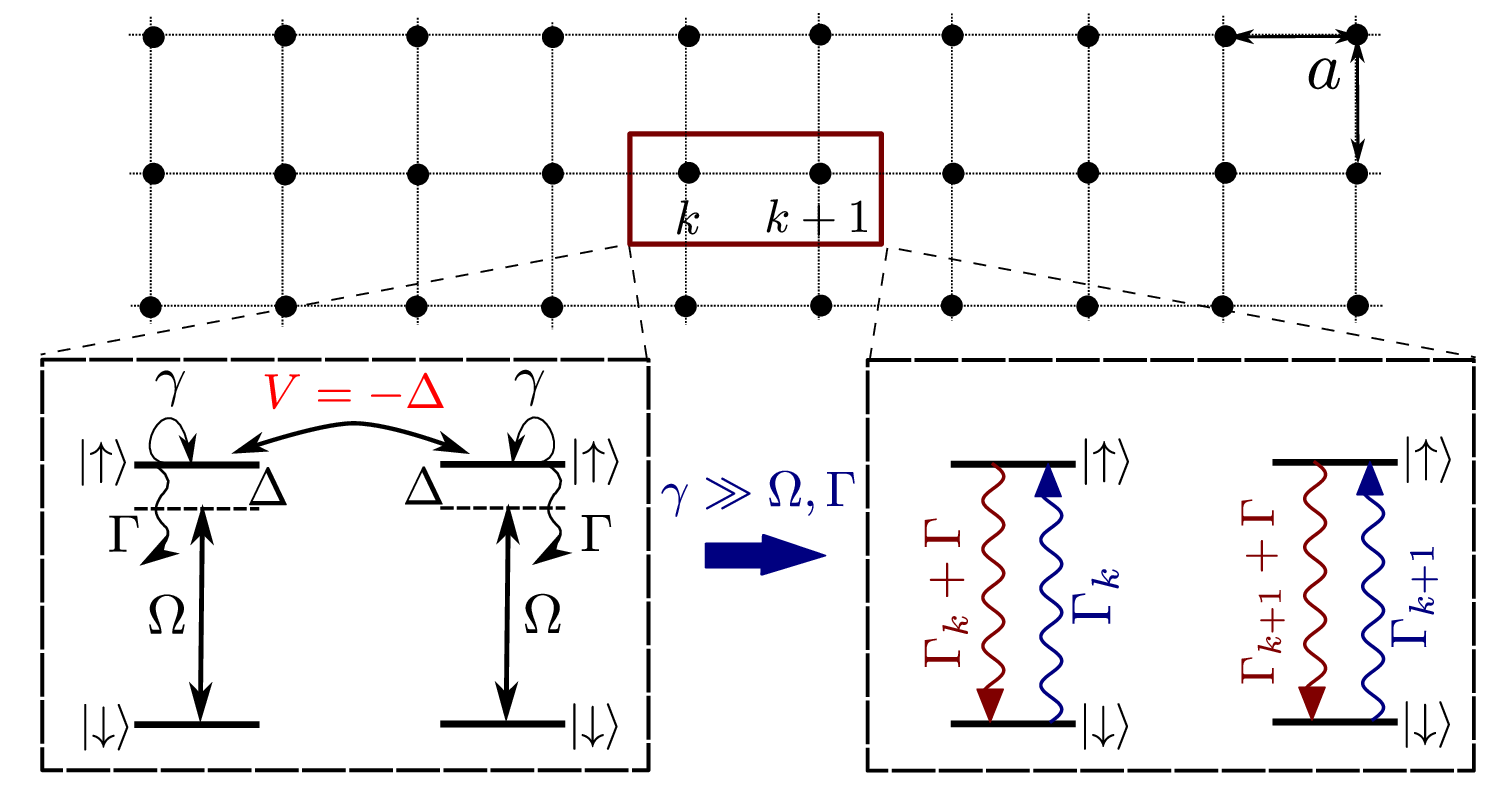}
  \caption{\textbf{Setup.} Two-level description of Rydberg atoms driven coherently by a laser field with Rabi frequency $\Omega$ and detuning $\Delta$. We also consider dephasing (at rate $\gamma$) of coherent superpositions between $\left|\uparrow\right>$ and $\left|\downarrow\right>$ and decay from $\left|\uparrow\right>$ to $\left|\downarrow\right>$ with rate $\Gamma$. Two atoms interact only when simultaneously excited to the Rydberg state $\left|\uparrow\right>$. The laser detuning $\Delta$ is chosen such that it exactly cancels out the nearest neighbour interaction $V$ ($\Delta = -V$). In the limit of strong dephasing the dynamics of this system can be described in terms of a classical stochastic process in which each spin flips with a rate $\Gamma_k$ that depends on the state of all atoms but the $k$-th one.}
\label{fig:fig1}
\end{figure}

The specific setup we consider (figure~\ref{fig:fig1}) is an ensemble of $N$ atoms trapped in a lattice with spacing $a$. Note that while we focus here on the one- and two-dimensional cases, the arguments below can be analogously applied to three dimensions as well. We describe the internal level structure of the atoms with two relevant levels: the ground state $\left|\downarrow\right>$ and a highly excited (Rydberg) state $\left|\uparrow\right>$ \cite{Rydberg2, Ryd-QI}, coupled by a laser with Rabi frequency $\Omega$ and detuning $\Delta$. When two atoms (at positions $\mathbf{r}_k$ and $\mathbf{r}_m$) are in the Rydberg state they experience a long range interaction of strength $V_{km}$ which is parameterized as $V_{km}=C_\alpha/\left|\mathbf{r}_k-\mathbf{r}_m\right|^\alpha$. We will focus here on Rydberg $s$-states that interact via van der Waals (vdW) forces ($\alpha=6$). The corresponding Hamiltonian of this system, expressed in a rotating frame and after discarding fast-oscillating terms (i.e. in the Rotating Wave Approximation) yields
\begin{equation}
  H=\Omega\sum_k\sigma_k^x+\Delta\sum_kn_k+\frac{1}{2}\sum_{k\neq m}V_{km}n_kn_m,
\end{equation}
where $n_k=\mathbb{1}-p_k=\left|\uparrow_k \right>\!\left<\uparrow_k  \right|$, $\sigma^k_x=\sigma_k^+ +\sigma_k^-$ and $\sigma_k^-=\left(\sigma_k^+\right)^ \dag=\left|\downarrow_k \right>\!\left<\uparrow_k \right|$. Moreover, we consider that the system is subject to two sources of Markovian noise (or dissipation), one that leads to spontaneous radiative decay from the Rydberg state to the ground state at a rate $\Gamma$ and the other causing dephasing of atomic superposition states at a rate $\gamma$ (due to uncorrelated noise) \cite{PRL-KinC,Marcuzzi14}. These two dissipation processes are described via the Lindblad superoperators
\begin{equation}
  {\cal L}_\Gamma \rho = \Gamma\sum_{k}\left( \sigma^-_k \rho \sigma^+_k - \frac{1}{2} \left\{n_k,\rho\right\}\right)
\end{equation}
and
\begin{equation}
  {\cal L}_\gamma\rho=\gamma\sum_k\left(n_k\rho n_k-\frac{1}{2}\left\{n_k,\rho\right\}\right),
\end{equation}
respectively.

In the limit of strong dephasing, i.e. $\gamma\gg\Gamma,\Omega$, a perturbative treatment of the full dissipative quantum dynamics allows for the description of the dynamics of this system by means of a classical rate equation for the probability vector $\mathbf{v}$ whose components are the statistical weights of the classical spin configurations (e.g., $\ket{\ldots \uparrow \uparrow \downarrow \ldots}$). The derivation of said equation is discussed in a more general framework in Refs. \cite{PRL-KinC,Marcuzzi14,Ates06, Ates07-2}, to which we refer the interested reader. Here we report the one corresponding to the setup outlined above:
\begin{equation}
	\dot{\mathbf{v}} = \sum_{k} \Gamma_k\left[\sigma_k^+-p_k\right]\mathbf{v} +  \sum_{k} \left(\Gamma+\Gamma_k\right)\left[\sigma_k^--n_k\right]\mathbf{v}.
	\label{eq:class}
\end{equation}
This depicts a classical stochastic process in which the $k$-th spin flips up with rate $\Gamma_k$ and down with rate $\Gamma+\Gamma_k$ (see figure \ref{fig:fig1}). Here, the rate
\begin{equation}
  \Gamma_k=\frac{\Omega^2\gamma}{\left( \frac{\gamma}{2} \right)^2 + \left(\Delta+\sum_{q\neq k}V_{kq}n_q\right)^2},
  \label{eq:rate}
\end{equation}
which depends on the state of all spins but the $k$-th one, is analogous to those that appear in dynamically-facilitated models of glasses \cite{Ritort03,Chandler10}. In \ref{app:qvscl} we compare the full quantum dissipative dynamics and the one described by (\ref{eq:class}) for small systems, thus providing further evidence of the validity of this approach in the current case.

\section{Emergent DP process}\label{sec:DP}

\begin{figure}
\center
  \includegraphics[width=0.8\columnwidth]{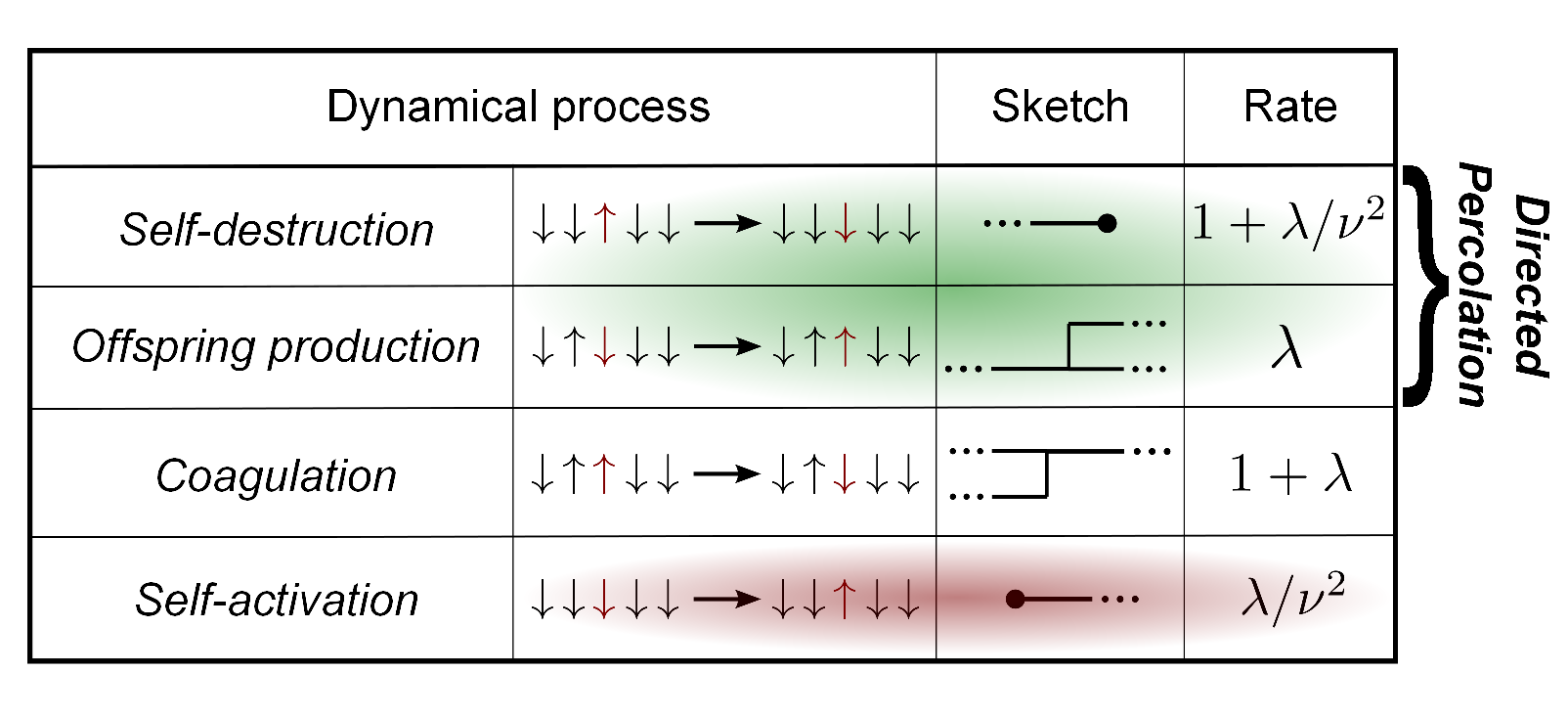}
  \caption{\textbf{Rates of the main dynamical processes.} Rates (in units of $\Gamma$) of the processes occurring in the Rydberg lattice (see main text). The self-destruction and offspring production (first and second row, shaded in green) of an excitation are fundamental processes of DP, while the creation of an isolated excitation or self-activation (fourth row, shaded in red) is a relevant perturbation away from the DP critical region. The remaining process (coagulation, third row) does not modify the critical properties.}
\label{fig:fig2}
\end{figure}

It has been conjectured that the defining conditions for the emergence of DP universality are: (i) a local dynamics with a unique absorbing state, (ii) a continuous phase transition with a positive, one-component order parameter and (iii) absence of additional symmetries \cite{DP_Janssen, DP_Hinrichsen}. In a spin chain, the simplest setup that meets these conditions consists of two fundamental processes: flipping down spins (self-destruction) and flipping them up provided there is an up spin in the neighbourhood (offspring production) (see two topmost rows in figure \ref{fig:fig2}). On the other hand, processes that can create isolated excitations (self-activation) destroy the absorbing property of the ``all-down'' state. Hence, they constitute a relevant perturbation away from the DP critical behaviour (fourth row in figure \ref{fig:fig2}). The presence of other processes, such as the flipping down of a spin provided there is another up spin in the neighbourhood (coagulation, third row in figure \ref{fig:fig2}) do not modify the critical properties of the system.

Let us now study the dynamical processes in our physical system and the corresponding underlying rates in more detail. While the self-destruction of excitations is simply provided by the radiative decay (rate $\Gamma$), the facilitated excitation and de-excitation of the $k$-th spin (offspring production and coagulation) occur at rates $\Gamma_k$ and $\Gamma+\Gamma_k$, respectively (see equation \reff{eq:rate}). These processes can be favoured by tuning the laser parameters adequately. In particular, given an existent excitation, fixing the detuning such that it cancels exactly the nearest neighbour interaction $V=C_6/a^6$, i.e. $\Delta=-V$, effectively brings on resonance the excitation and de-excitation of the atoms sitting next to it \cite{Lesanovsky14,Urvoy14}, i.e., it maximizes the value of $\Gamma_k$.

Clearly, due to the long-distance tails of the potential $V_{km}$, the presence of an excitation not only affects its neighbours, but slightly modifies the rates of spins lying far from it as well. As we argue in \ref{app:rates}, these changes are akin to introducing the possibility of L{\'e}vy flights in the offspring production. However, the van der Waals potential decays fast enough for the resulting corrections not to affect the critical properties of the DP dynamics \cite{Janssen1999, Hinrichsen2007}. We also refer to \ref{app:rates} for the derivation of the main dynamical processes of the Rydberg system. Here we report them in figure \ref{fig:fig2} along with the corresponding rates, re-expressed for simplicity in terms of the dimensionless parameters $\lambda = 4\Omega^2 / \gamma \Gamma$ and $\nu = 2 \abs{\Delta} / \gamma$ and measured in units of the decay rate $\Gamma$. Focussing on the fourth row, we note that the parameter $\nu$ can be increased (limit of large detuning $\abs{\Delta}=V\gg\gamma$) to reduce the impact of self-activation on the system. By comparing the first two rows, instead, we note that within the same regime $\lambda$ measures the relative strength of the offspring production and self-destruction processes and hence drives the DP phase transition. However, here the presence of long range interactions imposes an additional constraint on the actual emergence of DP universality, as it affects the growth of clusters of excitations. For illustration purposes, let us consider only the presence of next-to-nearest neighbour interaction of strength $\eta V$ with $\eta<1$: While the offspring production rate (i.e. $\downarrow\uparrow\downarrow\downarrow\downarrow \,\to \,\downarrow\uparrow\uparrow\downarrow\downarrow$) is given by $\lambda$, processes of the type $\uparrow\uparrow\downarrow\downarrow\downarrow \,\to \,\uparrow\uparrow\uparrow\downarrow\downarrow$ occur at a rate $\lambda/\left[1+\eta^2\nu^2\right]$. This can be seen calculating the corresponding energy difference $\Delta+\sum_{q\neq k}V_{kq}n_q=\Delta+V+\eta V=\eta V$ appearing at the denominator of (\ref{eq:rate}) for such a process. The latter rate, always smaller than $\lambda$, becomes extremely small in the limit of large $\nu$, eventually hindering the growth of clusters. To prevent this one needs to impose the constraint $\gamma \gtrsim \eta \abs{\Delta}$ ($ \eta\nu \lesssim 1$). In 1D and for a vdW potential we have $\eta= 1 / 2^6 = 1/64$ which allows one to make $\nu$ reasonably large. In contrast, in higher dimensions we find the more restrictive $\eta=1/8$, as next-to-nearest neighbours lie at a distance $\sqrt{2} a$ from one another. This restriction can be softened, as we discuss in \ref{app:potential}, by modulating the interaction between the atoms via external microwave fields.

\section{Mean-field analysis}\label{sec:MF}

As a simple consistency check, our first step towards understanding the critical properties of this non-equilibrium system is a mean-field analysis. The time evolution of the average local density of excitations $\left<n_k\right>$, after factorizing all spatial correlations $\left<n_kn_p\right>= \left<n_k\right> \left<n_p\right>$ ($k\neq p$), is governed by the equation $\partial_t \left<n_k\right> =  -\Gamma \left<n_k\right> + \left< 1 - 2n_k\right>\left<\Gamma_k\right>$. Employing the representation \reff{eq:projrate} and assuming the density to be uniform, i.e., $\av{n_k} = n \,\, \forall k$, one finds the mean-field equation
\begin{eqnarray}
	\partial_\tau n &=& - n +  \lambda (1-2n) \Bigl\{   z n(1-n)^{z-1}  \Bigr.\label{eq:nn1} \\ \nonumber
&& \Bigl.+ \sum_{j=0,2}^z\frac{1}{1+\nu^2\left|j-1\right|^2} {z \choose z-j}n^j(1-n)^{z-j}   \Bigr\},
\end{eqnarray}
where for convenience we have defined $\tau = \Gamma t$, and $z$ is the coordination number (i.e., number of nearest neighbours) of the lattice. Here, one can recognize in the first two terms spontaneous decay (self-destruction) and processes occurring where there is a single excitation around (offspring production and coagulation) (three first rows in figure \ref{fig:fig2}). The $j=0$ term, $\left[\lambda/\left(1+\nu^2\right)\right](1-2n)(1-n)^z$, represents the self-activation process, which is the one driving the system away from criticality. This can be understood by noticing that it is the only surviving term when $n=0$, hence allowing the system to leave the ``all-down'' state.

\begin{figure}
\center
  \includegraphics[width=0.7\columnwidth]{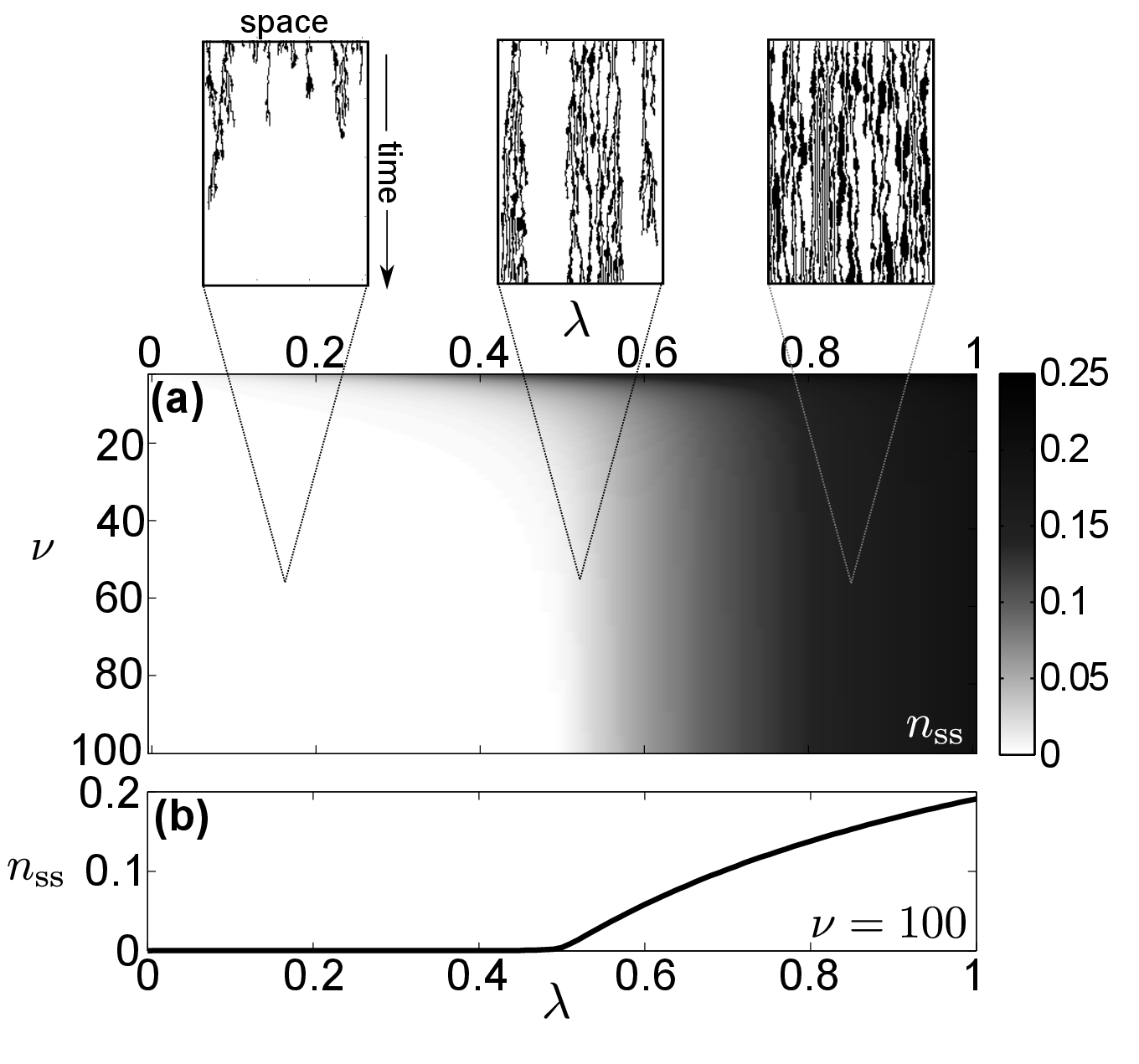}
  \caption{\textbf{Mean-field phase diagram.} \textbf{(a)} Mean-field stationary state density $n_\mathrm{ss}$ as a function of $\lambda$ and $\nu$ (see main text) with $z=2$. The insets display typical realizations of the time evolution of a one-dimensional chain of spins for three different values of $\lambda$. Black dots stand for $\ket{\uparrow}$, whereas white ones for $\ket{\downarrow}$. \textbf{(b)} Corresponding profile of $n_\mathrm{ss}$ as a function of $\lambda$ at fixed $\nu = 100$.}
\label{fig:fig3}
\end{figure}

The stationary mean-field solution as a function of $\lambda$ and $\nu$ is shown in figure \ref{fig:fig3}(a) for a 1D chain ($z=2$). For finite $\nu$ we observe a smooth crossover from a low-density region to a high-density one. As we increase $\nu\gg 1, \lambda$ [see figure \ref{fig:fig3}(b)], criticality emerges and one observes a sharp transition at $\lambda_c=1/2$. In this regime, the last terms of \reff{eq:nn1} affect the evolution only on long timescales --- dictated for a single spin by $\nu^2 / (\lambda \Gamma) = \Delta^2/(\Omega^2\gamma)$. The dynamics at times shorter than those is thus effectively captured by
\begin{equation}
	\partial_\tau n = - n +  \lambda z n (1-2n) (1-n)^{z-1},
	\label{eq:EEOM2}
\end{equation}
which is similar to the mean-field equations of other known DP processes \cite{DP_Hinrichsen}. In fact, it features a DP-like phase transition: while for $\lambda z<1$ the only stationary state is the absorbing one with density $n_\mathrm{ss}=0$, for $\lambda z > 1$ the latter becomes unstable and a new, stable one with a non-vanishing density appears, identifying $\lambda_c = 1/z$ as a critical point.

The mean-field critical exponents can also be extracted: the approach to stationarity can be studied at the critical point by introducing a small perturbation from the all-down state and linearising \reff{eq:EEOM2}, which yields $\partial_\tau n = - (z+1) n^2$ and therefore agrees with the prediction for DP $n \sim t^{-\delta}$ with $\delta^{(\mathrm{MF})} = 1$. Focusing on the properties of the steady state as a function of $\lambda$ in the proximity of the critical point ($\lambda z  = 1+ \epsilon$) one finds, at leading order, $ n_\mathrm{ss} \approx \epsilon / (z+1)$, which is again compatible with the DP prediction $n_\mathrm{ss} \sim \epsilon^\beta$ with $\beta^{(\mathrm{MF})} = 1$ \cite{DP_Hinrichsen}.

\section{Numerical results}\label{sec:Numerics}

\begin{figure}
\center
  \includegraphics[width=0.8\columnwidth]{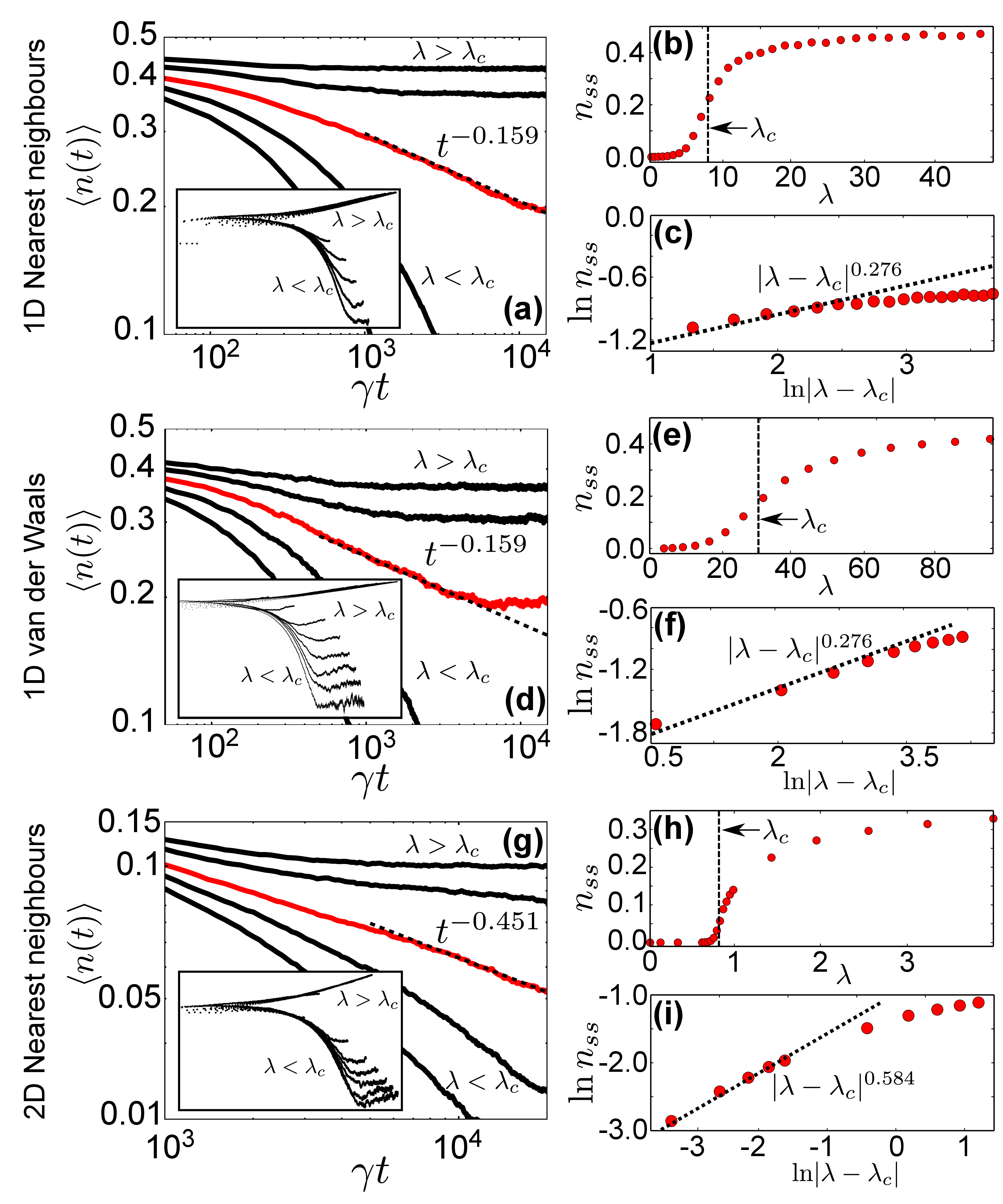}
  \caption{\textbf{Numerical simulations.} \textbf{(a)} Time evolution of the density of excitations for a 1D system with $50$ spins with NN interactions and $\nu = 128$. Each of the five lines correspond to a different value of $\lambda$. Both the upper and the lower ones show exponential relaxation towards the respective stationary values. The central one, highlighted in red, features instead an approximate algebraic decay with a slope close to the one expected for DP (dashed line, shown for comparison). The inset shows the collapse of these curves after assuming validity of DP scaling (see main text). \textbf{(b)} Profile of the stationary density (at $t = 1.5\cdot10^4 \gamma^{-1}$) as a function of $\lambda$ for NN, showing a crossover from a low-density to a high-density phase. \textbf{(c)} Using the data in (a) to estimate the position of the critical point $\lambda_\mathrm{c}$ we highlight an algebraic behaviour which recalls the DP one (dotted line, shown for comparison). \textbf{(d-e-f)} Analogous plots for a 1D system of $50$ spins with vdW interactions. The dynamical profiles of the excitation density still collapse onto two distinct master curves, although the disrupting effects of self-activation are effectively amplified due to the difficulty of growing large clusters, as discussed in the main text. \textbf{(g-h-i)} Analogous plots for a 2D system of $20 \times 20$ spins with NN interactions.}
\label{fig:1d2d}
\end{figure}

We have performed Monte Carlo dynamic simulations of small 1D and 2D systems with open boundary conditions for \reff{eq:class} (see figure \ref{fig:1d2d}). While \reff{eq:class} can be simulated for very large lattices, we focus here in system sizes and boundary conditions relevant to current Rydberg experiments. We have used $\nu = 128$ and the initial state is a random classical spin configuration with fixed global density $\left<n(0)\right> = \sum_k \av{n_k (0)}/N =  0.4$.

We first focus on the 1D case, where we have accounted for both vdW ($\eta = 1/64$) and nearest neighbour (NN) interactions ($\eta=0$) in a system of $N=50$ atoms. Signatures of the transition are most effectively highlighted in the dynamics. By increasing $\lambda$ (e.g., by acting on the Rabi frequency $\Omega$), one encounters a qualitative change of the curves from an exponential decay towards $0$ (lower curves) to an exponential relaxation to a finite value (upper curves) with a characteristic bending down and up, respectively [see figures \ref{fig:1d2d}(a) and (d)]. The intermediate, critical regime ($\lambda=\lambda_\mathrm{c}$) displays instead an algebraic decay in reasonable agreement with the DP one \cite{DP_Hinrichsen} ($t^{-\delta}$ with $\delta^{(1D)}=0.159$ is shown for comparison). We then analyze the behaviour of the stationary density in figures \ref{fig:1d2d}(b) and (e). This yields again an algebraic law close to $\lambda_\mathrm{c}$ which resembles the DP one [the expected value $\beta^{(1D)} = 0.276$ is again shown for comparison in panels (c) and (f)]. As a further check, we have assumed scaling with the DP critical exponents \cite{DP_Hinrichsen} and plotted $t^{\delta^{(1D)}} \left<n(t)\right>$ as a function of $t|\lambda-\lambda_\mathrm{c}|^{\beta^{(1D)}/\delta^{(1D)}}$ for different values of $\lambda$ close to the critical point as an inset in panels (a) and (d). This shows that the curves tend to collapse, as expected in the presence of a continuous phase transition. This collapse is not present at short times, since we have not accounted for initial slip effects, and at long times $t\gg\Delta^2/(\Omega^2\gamma)$, where the relevant processes perturbing DP become non-negligible. As discussed earlier, the main difference between the dynamics resulting from the two interactions (NN and vdW) is the rate of growth from a two- to a three-excitation cluster ($\uparrow\uparrow\downarrow\downarrow\downarrow \,\to \, \uparrow\uparrow\uparrow\downarrow\downarrow$) which, with our choice of $\nu = 128$, is $5$ times smaller in the vdW case. Correspondingly, we expect a shift in the critical point by approximately the same factor, as can be indeed observed by comparing figures \ref{fig:1d2d}(b) and \ref{fig:1d2d}(e). Note that the profiles of these curves are analogously stretched by the same factor, making the nearest-neighbour crossover look sharper than the vdW one. We have verified, however, that the two curves plotted as functions of the rescaled parameter $(\lambda - \lambda_c)/\lambda_c$ look reasonably similar. Moreover, as the value of $\lambda_\mathrm{c}$ is increased in the vdW case, the time at which the self-activation mechanism becomes relevant (proportional to $\nu^2 / (\lambda \Gamma)$) is shorter. Despite this, one can still observe a collapse of the curves in the intermediate regime as discussed above.

The dynamics in 2D with vdW interactions is actually more complicated: at the rather small sizes accessible in experiments and numerically investigated here, one observes the emergence of a spurious one-dimensional, rather than two-dimensional critical behaviour. This can be intuitively understood by analyzing once again the growth from a two- to a three-excitation cluster: The two neighbouring excitations identify a direction (say, along a row or a column of the square lattice). Further growth along this direction occurs at a rate $\lambda/\left[1+(\nu/64)^2\right]$.
In contrast, the rate for growing in the orthogonal direction is $\lambda/\left[1+(\nu/8)^2\right]$, since in this case the next-to-nearest excitation lies at distance $\sqrt{2}a$ from the first excitation (i.e., along the diagonal of a square cell). Hence, it is much more likely to grow linear structures than it is to percolate in both directions. At the same time, this anisotropy is not a geometrical property of the system, but it is stochastically determined for a cluster by the first excitation it forms. In small systems, clusters growing vertically and horizontally hinder the respective growths, effectively reducing the available amount of space and thus, in a sense, amplifying finite-size effects. In order to recover proper 2D directed percolation behaviour in this setting one would have to increase the system size beyond the current experimental possibilities. We remark that, although lowering the detuning reduces the effective anisotropy, it also increases the self-activation rate and is therefore not a valid strategy. A seemingly more viable approach is to shape the atomic interaction potential.
In \ref{app:potential} we discuss a way to modulate the potential and induce a faster decay of the tail at large distances. This allows to employ the nearest-neighbour case as an approximate description in higher dimensions. The 2D data shown in figures \ref{fig:1d2d}(g), (h) and (i) are therefore obtained for a system of $N=20\times20$ atoms considering NN interactions only. They show that the exponents $\delta^{(2D)}=0.451$ and $\beta^{(2D)}=0.584$ fit again fairly well the data from the numerical simulations. Accordingly, the collapse also displays the emergence of two distinct master curves.

\section{Conclusions}\label{sec:Conclusions}

We have shown that interacting gases of Rydberg atoms feature a microscopic dynamics that leads to an emergent out-of-equilibrium behaviour displaying specific features characteristic of DP. Surprisingly this is the case already for rather small mesoscopic system sizes which are currently studied experimentally. This renders these cold atomic systems into a viable candidate for the experimental observation of the simple --- yet intriguingly hard to observe --- DP non-equilibrium universality class. Such experimental studies could furthermore become a platform for a systematic exploration of the role of quantum effects on the dynamics (a very challenging task from the theoretical point of view), and shed light on the role of the assumed ordered configuration (lattice) by considering continuous gases.

\ack
The authors thank S. Diehl for fruitful discussions. The research leading to these results has received funding from the European Research Council under the European Union's Seventh Framework Programme (FP/2007-2013) / ERC Grant Agreement No. 335266 (ESCQUMA), the EU-FET grant HAIRS 612862 and from the University of Nottingham. Further funding was received through the H2020-FETPROACT-2014 grant No.  640378 (RYSQ). We also acknowledge financial support from EPSRC Grant no.\ EP/J009776/1. B.O. acknowledges funding from the Royal Society.

\appendix

\section{Effectively classical dynamics}\label{app:qvscl}

\begin{figure}
\center
\includegraphics[width=0.7\columnwidth]{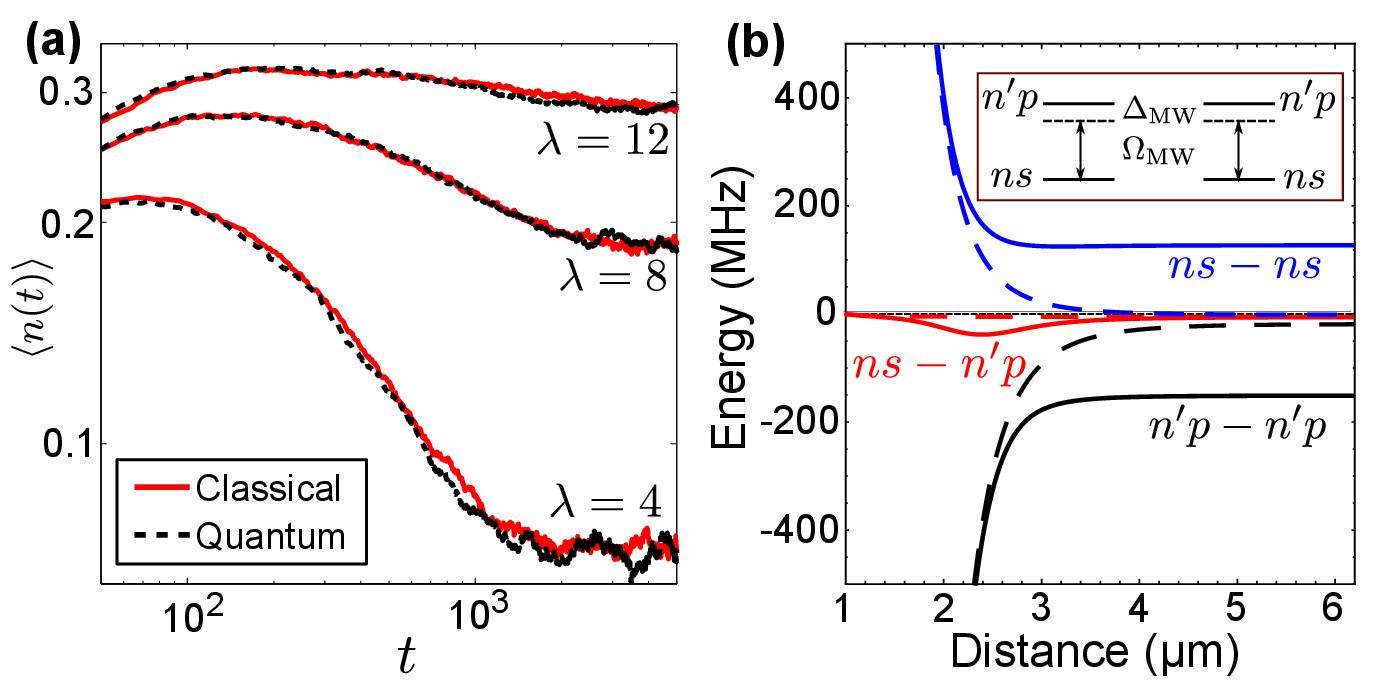}
\caption{\textbf{Potential shaping and effectively classical dynamics.} \textbf{(a)} Time evolution of the density of excitations starting from a single excitation in the center of a 1D chain. We compare the results obtained from the simulation of the full quantum Master equation via Quantum Jump Monte Carlo ($2\times10^3$ trajectories) and the effective classical equation (\ref{eq:class}) via Monte Carlo ($10^4$ trajectories) for $N=9$ atoms. The parameters used in the simulations shown are $\Delta=-V=-10\gamma$, $\Omega/\gamma=0.1$ and $\Gamma/\gamma=1/100,1/200$ and $1/300$, i.e. $\lambda=4,8$ and $12$. \textbf{(b)} Potential shaping via microwave dressing of the state $52s$ state with state $57p$ (rubidium). The dashed lines show the interactions between the bare states while the solid ones show the dressed potential. }\label{fig:qvscl}
\end{figure}

Here, we provide evidence of the validity of the effectively classical equation of motion \reff{eq:class} in the specific parameter regime considered in this work. A numerically exact solution of the full quantum Master equation underlying this problem is only possible for small systems sizes. Here we use $N=9$ atoms in a 1D chain with NN interactions. Initially there is a single excitation in the center of the lattice and the corresponding dissipative quantum dynamics is simulated with Quantum Jump Monte Carlo. A comparison to the dynamics obtained from \reff{eq:class} is displayed in figure \ref{fig:qvscl}(a) for $\Delta=-V=-10\gamma$, $\Omega/\gamma=0.1$ and $\Gamma/\gamma=1/100,1/200$ and $1/300$. Indeed an excellent agreement between the full quantum and the approximate classical dynamics is found \cite{Marcuzzi14}.

\section{Adjusting the rates of dynamical processes in the Rydberg lattice}\label{app:rates}

We introduce the projectors $\mal{P}^{(j)}_k$ ($j=1\dots z$, with $z$ being the coordination number of the lattice) on the subspaces where exactly $j$ nearest neighbours of the $k$-th atom are excited. As these projectors commute with $\Gamma_k$, we can decompose $\Gamma_k = \Gamma_k \sum_j \mal{P}^{(j)}_k  = \sum_j \Gamma_k^{(j)}$ with $\Gamma_k^{(j)} =\mal{P}^{(j)}_k \Gamma_k \mal{P}^{(j)}_k$, which yields
\begin{equation}
	 \Gamma_k^{(j)} = {\cal P}^{(j)}_k \frac{\Omega^2\gamma}{\left(\frac{\gamma}{2}\right)^2  +  \left[\Delta (1-j) + \sum_{q \neq k, \left\{ \bar{k} \right\}  }  V_{kq} n_q  \right]^2}  {\cal P}^{(j)}_k,
	 \label{eq:projrate}
\end{equation}
where $\left\{ \bar{k}  \right\}$ denotes all sites included in the neighbourhood of site $k$. Consequently, $q \neq k, \left\{ \bar{k} \right\} $ restricts the sum on atoms lying further than nearest neighbours from site $k$. Since the interaction is quickly decaying with the distance we have  $\sum_{q \neq k, \left\{ \bar{k} \right\}} V_{qk} \ll \gamma$ which allows us to approximate $\Gamma_k^{(1)} \approx (4\Omega^2 / \gamma) \mal{P}_k^{(1)}$. This represents the rate at which a spin flips up if it has a single excited neighbour (second row in figure \ref{fig:fig2}). It is important to remark that this approximation, made at the microscopic scale, is expected not to influence the critical behaviour emerging on mesoscopic scales: in fact, the leading correction can be expressed as
\begin{equation}
	\frac{4\Omega^2 }{ \gamma} -  \frac{\Omega^2\gamma}{\left(\frac{\gamma}{2}\right)^2  +  \lt \sum_{q \neq k, \left\{ \bar{k} \right\}  }  V_{kq} n_q  \rt^2} \approx \frac{4\Omega^2 }{ \gamma} \lt \frac{4}{\gamma} \sum_{q \neq k, \left\{ \bar{k} \right\}  }  V_{kq} n_q  \rt^2.
	 \label{eq:ratecorr}
\end{equation}
Exploiting the fact that the operator inside the brackets is diagonal in the $\sigma^z$ basis, we see that for every possible classical state it can be bounded from above according to
\begin{equation}
	\frac{4\Omega^2 }{ \gamma} \lt \frac{4}{\gamma} \sum_{q \neq k, \left\{ \bar{k} \right\}  }  V_{kq} n_q  \rt^2 \lesssim    \frac{64\Omega^2 }{ \gamma^3 }   \lt \sum_{q \neq k, \left\{ \bar{k} \right\}  }  V_{kq}  \rt     \lt  \sum_{q \neq k, \left\{ \bar{k} \right\}  }  V_{kq} n_q \rt
	\label{eq:ratecorr2}
\end{equation}
which is reminiscent of a term causing offspring production at arbitrary distances from the parent excitation with an algebraically-decreasing rate. Such corrections have been investigated in the past \cite{Janssen1999, Hinrichsen2007} and have been found to modify the scaling behaviour only if the exponent of the aforementioned algebraic decay lies below a certain threshold. The latter was estimated to be $2.077$ in 1D and $4.2$ in 2D (see, e.g., equation (42) in \cite{Hinrichsen2007}). For van der Waals interactions, the exponent appearing in equation \reff{eq:ratecorr2} is $6$ and thus definitely higher than these thresholds and hence the dynamics is deep within the ``ordinary'' DP regime. 

Correspondingly, $\Gamma_k^{(0)}$ provides an estimate of the rate at which spins flip up when far away from excitations. These processes can create isolated excitations, thereby destroying the absorbing property of the ``all-down'' state. They therefore constitute a relevant perturbation away from the DP critical region (fourth row in figure \ref{fig:fig2}). Using the fact that the Rydberg interaction is quickly decaying, i.e. $ \sum_{q \neq k, \left\{ \bar{k} \right\}} V_{qk} \ll V = \abs{\Delta} $, the rate of these processes can be estimated as
\begin{equation}
	 \Gamma_k^{(j \neq 1)}   \approx   \frac{4\Omega^2}{\gamma}\left[  \frac{1}{1  +  \left(\frac{2\Delta}{\gamma}\right)^2 |j-1|^2}\right]        {\cal P}^{(j \neq 1)}_k.
	 \label{eq:Gapprox}
\end{equation}
Hence, the magnitude of the ``DP-breaking'' processes is strongly suppressed at large laser detuning: when $|\Delta| \gg \gamma$ one finds that $	\|\Gamma_k^{(j \neq 1)} \|  \leq  \| \Gamma_k^{(0)} \| \approx   \frac{\Omega^2\gamma}{\Delta^2} \ll \frac{4\Omega^2}{ \gamma} \approx \|\Gamma_k^{(1)}\|$ where $\|A\|$ denotes the largest eigenvalue of $A$ in modulus and, in the simple case of \reff{eq:Gapprox}, it can be straightforwardly obtained by replacing the projectors ${\cal P}$ with $1$.

\section{Potential shaping}\label{app:potential}

The Rydberg states we are considering here display long range interactions of the vdW type. As we have discussed in the main text, in one dimension they allow for the observation of DP in 1D. In 2D and 3D, however, the presence of long range interactions induces critical properties different from the 2D and 3D DP ones due to the preference of the clusters to grow linearly. In order to observe DP in Rydberg gases in dimensions two and three, it is desirable to have a potential closer to a NN interaction.

This can be achieved through potential shaping via microwave dressing of Rydberg states \cite{Petrosyan14,Kiffner13,Kiffner13-1}. To illustrate this we consider a Rydberg $s$-state (orbital quantum number $l=0$) with principal quantum number $n$ coupled to a $n'p$ state via a linearly polarized microwave field with Rabi frequency $\Omega_\mathrm{MW}$ and detuning $\Delta_\mathrm{MW}$ [see figure \ref{fig:qvscl}(b)]. This results in a ``hybridization'' of the respective interactions, which consequently modifies the bare $ns-ns$ purely algebraic decay of vdW potential. For example, in figure \ref{fig:qvscl}(b) we show the result of coupling the $52s$ and $57p$ states of Rubidium with Rabi frequency $\Omega_\mathrm{MW}=70$ MHz and detuning $\Delta_\mathrm{MW}=-10$ MHz. Considering the lattice constant to be $a=2\,\mu m$, the ratio between nearest (distance $a$) and next-to-nearest neighbour ($\sqrt{2}a$ in 2D and 3D) interaction increases from $\eta^{-1}= 8 $ (in the vdW case) to $\eta^{-1} \approx 51$.

Note that for the coherent superposition between the two states ($ns$ and $n'p$) to be maintained throughout the experiment the Rabi frequency of the microwave field must be the largest energy scale in the problem, i.e. $\Omega_\mathrm{MW}\gg\gamma\gg\Gamma,\Omega$.

\section*{References}
\providecommand{\newblock}{}

\end{document}